\documentclass[reprint,eqsecnum,floats,aps,amsmath,amssymb,nofootinbib,prd,onecolumn, showpacs]{revtex4-2}
\usepackage{graphicx,physics}
\usepackage{amsmath,amssymb,mathtools,mathrsfs}
\usepackage{hyperref}
\usepackage{graphicx}
\usepackage{subfigure}
\usepackage{arydshln}
\usepackage{xcolor}
\usepackage{braket}
\usepackage{tensor}
\usepackage{enumitem,array,textcomp}
\usepackage[utf8x]{inputenc}

\DeclareMathOperator\arctanh{arctanh}

\begin{document}

\title{\textbf{Space of solutions of the Ashtekar-Olmedo-Singh effective black hole model}}
\author{Beatriz Elizaga Navascués}
\email{w.iac20060@kurenai.waseda.jp}
\affiliation{JSPS International Research Fellow, Department of Physics, Waseda University, 3-4-1 Okubo, Shinjuku-ku, 169-8555 Tokyo, Japan}

\author{Alejandro Garc\'ia-Quismondo}
\email{alejandro.garcia@iem.cfmac.csic.es}
\affiliation{Instituto de Estructura de la Materia, IEM-CSIC, Serrano 121, 28006 Madrid, Spain}

\author{Guillermo  A. Mena Marug\'an}
\email{mena@iem.cfmac.csic.es}
\affiliation{Instituto de Estructura de la Materia, IEM-CSIC, Serrano 121, 28006 Madrid, Spain}

\begin{abstract}
We consider a general choice of integration constants in the resolution of the dynamical equations derived from a recently proposed effective model that describes black hole spacetimes in the context of loop quantum cosmology. The interest of our analysis is twofold. On the one hand, it allows for a study of the entire space of solutions of the model, which is absent in the literature and is fundamental for understanding the relation with any underlying quantum theory. On the other hand, choices of integration constants that generalize the type of solutions considered so far may lead to exotic behaviors in the effective black hole geometry, as well as modified thermodynamical properties. With these motivations in mind, we discuss the interior and exterior geometries, and present the conditions that a satisfactory matching at the horizons imposes. Then, we turn our attention to the Hawking temperature associated with the black horizon of the model, which we find to be affected by the freedom of choice of integration constants. Finally, we briefly comment on the asymptotic structure of the general solution and compare different notions of mass.  
\end{abstract}

	
\maketitle

%%%%
\section{Introduction}
 
Ashtekar, Olmedo, and Singh (AOS) \cite{AOS,AOS2} have put forward a new effective model that aims to describe non-rotating, uncharged black holes within the context of loop quantum cosmology (LQC) \cite{ALQG,Thiem,AS,LQCG}. This model stands out among previous related works (see e.g. Refs. \cite{1,2,3,18b,4,5,6,7,8,9,10,11,12,13,14,15,16,19,20,21,22,23}) for a number of reasons. On the one hand, it is claimed to be free from pathological properties such as the dependence on fiducial structures and the appearance of local quantum effects in regions of low spacetime curvature. On the other hand, it leads to an effective quantum extension of the entire Kruskal spacetime where the curvature invariants remain finite, even across the spatial hypersurface that replaces the classical central singularity, which is interpreted to mediate a transition between a trapped region and an anti-trapped one. Notwithstanding its appealing features, the model has received a certain degree of criticism, mainly concerning possible problems with its asymptotic structure \cite{Bouhm,AO}, covariance properties \cite{bojosb}, and the relation between the proposed effective Hamiltonian and the studied dynamics \cite{N}. This last point is intimately related to the fact that the polymerization parameters (which regulate the introduction of quantum effects in the system, in the sense that the effective Hamiltonian reduces to that of general relativity when the parameters vanish), while claimed to be Dirac constants of motion, are nonetheless treated as constant numbers in the derivation of the Hamiltonian equations of motion. This strategy was supported in Ref. \cite{AOS2} by arguments based on an extension of phase space. Alternatively, in order to try and reconcile the physical results of the AOS model with a more standard treatment of the polymerization parameters as constants of motion, a two-time description has been recently proposed \cite{AG,AG2}.

Since this effective model is attracting a considerable attention owing to its good physical properties (in spite of the commented caveats), a next logical step would be to proceed to construct a quantum version along the lines of LQC. It is well known that there is a close relation between quantizing a system and characterizing its space of solutions. In the case at hand, this requires an understanding of the full extent of the dynamics of the AOS model. In the original works where this model was discussed \cite{AOS,AOS2}, the constants that arise in the integration of the dynamical equations were fixed in a particular way such that one recovers the classical solutions of general relativity when the polymerization parameters vanish. This choice confines the analysis to a subspace of the space of solutions of the model. However, a careful inspection proves that requiring a satisfactory classical behavior may at most fix the dominant part of the integration constants in a suitable asymptotic regime. This observation opens the possibility of a more general choice of integration constants while retaining a sensible classical limit. Such a less restrictive choice is interesting not only from the point of view of exploring the full space of admissible solutions of the AOS dynamical equations, but also because deviations from the previously considered values may leave a physical imprint, e.g. in the thermodynamical properties of the black hole or in the form of the shadow cast by it (see Ref. \cite{Shadow} for an analysis of the shadow of an AOS black hole, which does not show important quantum effects for large black hole masses in Planck units). Thus, the motivation to consider a general choice of integration constants is twofold. On the one hand, it allows us to analyze the whole space of solutions of the AOS model, which is of a fundamental importance if we wish to carry out its quantization. On the other hand, it clarifies the physical role of the integration constants and how they affect, for instance, the thermodynamical properties or the different notions of mass for the spacetime. Throughout this paper, we focus our study exclusively on the AOS equations of motion and their associated space of solutions, bearing in mind that our aim is to set a groundwork that facilitates the quantization of the model in future works.

The present paper is structured as follows. In Sec. \ref{sec:GeneralSolution}, we write down the dynamical equations and solve them for a general choice of integration constants. This analysis is performed both in the interior and exterior regions (see Secs. \ref{subsec:Interior} and \ref{subsec:Exterior}, respectively), showing that an acceptable matching of solutions at the black and white horizons is possible. Then, we investigate the thermodynamics of the general solution in Sec. \ref{sec:Thermodynamics}, putting a special emphasis on the Hawking temperature and the relation between the analog of the Schwarzschild radius and the Hamiltonian mass. In Sec. \ref{sec:AsympMass}, we comment on the behavior of the general solution at spatial infinity and relate the Hamiltonian mass with other definitions of the mass directly associated with the black hole spacetime. Finally, in Sec. \ref{sec:Conclusions}, we summarize the main results of this work and discuss their consequences. Throughout this article, we adopt natural units, setting the speed of light and the reduced Planck constant to one.   

%%%%
\section{General solution of the AOS dynamical equations}\label{sec:GeneralSolution}

Let us derive the general solution of the equations of motion obtained in Refs. \cite{AOS,AOS2}. We will address the case of the interior region in Sec. \ref{subsec:Interior}, and study the exterior region and the matching at the horizons in Sec. \ref{subsec:Exterior}.

\subsection{Interior region}\label{subsec:Interior}

We consider the interior region of a non-rotating, uncharged black hole, which can be foliated by homogeneous but anisotropic spacelike Cauchy hypersurfaces. We can formulate a Hamiltonian description of the relativistic system on a finite dimensional phase space. Following the standard procedure in LQC, the dynamical degrees of freedom are captured in a gravitational $\mathfrak{su}(2)$ connection and a densitized triad. Once the symmetries of the system are taken into account, all the dynamical information contained in these fields is encoded in a total of four basic variables: the so-called connection variables $b$ and $c$, and the triad variables $p_b$ and $p_c$. They form two canonical pairs, $\{b,p_b\}$ and $\{c,p_c\}$. In the AOS model, their dynamics is governed by the following equations \cite{AOS2}:
\begin{align}
c'&=-2\dfrac{\sin(\delta_cc)}{\delta_c},\label{c'}\\
p_c'&=2p_c\cos(\delta_cc),\label{pc'}\\
b'&=-\dfrac{1}{2}\left[\dfrac{\sin(\delta_bb)}{\delta_b}+\dfrac{\gamma^2\delta_b}{\sin(\delta_bb)}\right],\label{b'}\\
p_b'&=\dfrac{1}{2}p_b \cos(\delta_bb)\left[1-\dfrac{\gamma^2\delta_b^2}{\sin^2(\delta_bb)}\right],\label{pb'}
\end{align}
where the prime denotes the derivative with respect to the coordinate time $t$, $\gamma$ is the Immirzi parameter \cite{Immirzi}, and $\delta_b$ and $\delta_c$ are the polymerization parameters that control the inclusion of quantum effects in the system (indeed, the classical, general relativistic equations of motion are recovered if both parameters tend to zero). Originally, these dynamical equations were motivated starting with an effective Hamiltonian. With the choice of lapse function $N=\gamma \delta_b\sqrt{|p_c|}/\sin(\delta_bb)$, the product of the effective Hamiltonian constraint $H_{\rm eff}$ and the lapse can be written as \cite{AOS,AOS2}
\begin{align}
N H_{\rm eff}&=\dfrac{L_o}{G}(O_b-O_c),\qquad O_b=-\dfrac{1}{2\gamma}\left[\dfrac{\sin(\delta_bb)}{\delta_b}+\dfrac{\gamma^2\delta_b}{\sin(\delta_bb)}\right]\dfrac{p_b}{L_o},\qquad O_c=\dfrac{1}{\gamma}\dfrac{\sin(\delta_cc)}{\delta_c}\dfrac{p_c}{L_o}, \label{Heff}
\end{align}
where $L_o$ is a fiducial length related to the radial direction of the spatial sections and $G$ is the Newton gravitational constant. However, it is worth making clear that we concentrate here exclusively on the investigation of the space of solutions of Eqs. \eqref{c'}-\eqref{pb'}, regardless of the way in which they are obtained.

The considered dynamical equations can be solved in a straightforward manner following the same strategy as in Ref. \cite{AOS2}. The effective solutions resulting from the integration of Eqs. \eqref{c'} and \eqref{pc'} are
\begin{align}
\tan \left[\dfrac{\delta_cc(t)}{2}\right]=C_1 e^{-2t}=\textrm{sgn}(C_1)x_c(t),\qquad p_c(t)=\dfrac{1}{2}\bar p_c^0\left[x_c(t)+\dfrac{1}{x_c(t)}\right],
\end{align}
where $C_1$ and $\bar p_c^0$ are non-zero, real integration constants, and $x_c(t)=|C_1|e^{-2t}$ is strictly positive. Similarly, the solution to the equation of motion corresponding to $b$ \eqref{b'} reads
\begin{align}
\cos \left[\delta_bb(t)\right]=b_o\tanh \left(\dfrac{1}{2}b_o t+C_2\right),
\end{align}
where $b_o=\sqrt{1+\gamma^2\delta_b^2}$ and $C_2$ is a real integration constant. Finally, in order to obtain $p_b(t)$, it is not necessary to integrate Eq. \eqref{pb'}. Indeed, since the system under consideration is gravitational in nature, it inherits from general relativity the property that it is fully constrained. Once we have selected a spacetime foliation in homogeneous hypersurfaces and chosen our set of canonical variables, these are only subject to one non-trivial constraint, namely, that the effective Hamiltonian in Eq. \eqref{Heff} vanishes. Plugging the effective solutions derived so far into this constraint leads to
\begin{align}\label{pb}
p_b(t)=-2\dfrac{\delta_b}{\delta_c}\dfrac{\sin\left[\delta_bb(t)\right]}{\gamma^2\delta_b^2+\sin^2\left[\delta_bb(t)\right]}p_c(t)\sin\left[\delta_cc(t)\right]=-\dfrac{2\gamma L_o \delta_bm}{b_o^2}\dfrac{\sqrt{1-b_o^2\tanh^2\left[\dfrac{1}{2}b_ot+C_2\right]}}{1-\tanh^2\left[\dfrac{1}{2}b_o t+C_2\right]}\textrm{sgn}(\sin\left[\delta_bb(t)\right]),
\end{align}
where $m=\textrm{sgn}(C_1)\bar p_c^0 /(\gamma L_o \delta_c)$ is the coincident on-shell value of $O_b$ and $O_c$. The absolute value of this constant of motion is proportional to the ADM mass of the black hole in the AOS model (see Ref. \cite{AO}). We will call it the Hamiltonian mass and investigate its precise relation to the mass of the black hole for a general choice of integration constants in Sec. \ref{sec:AsympMass}.

Thus, the interior solution is determined by three real integration constants ($C_1$, $C_2$, and $\bar p_c^0$) and two positive polymerization parameters ($\delta_b$ and $\delta_c$) that are restricted to display a concrete dependence on $|m|$ for large black hole masses (see Ref. \cite{AOS2}). We have already seen that $\bar p_c^0$ is directly related with the Hamiltonian mass [namely $\bar p_c^0=\textrm{sgn}(C_1)\gamma L_o\delta_c m$]. Let us briefly comment on the interpretation of the two remaining integration constants, $C_1$ and $C_2$. In terms of the triad and connection variables, the effective spacetime metric can be written as
\begin{align}
ds^2_{\rm int}=-\dfrac{\gamma^2 \delta_b^2|p_c|}{\sin^2(\delta_bb)}dt^2+\dfrac{p_b^2}{L_o^2|p_c|}dx^2+|p_c|d\Omega^2,\label{metric}
\end{align}
where we have employed the choice of lapse function $N$ given above Eq. \eqref{Heff}, $x$ is a radial coordinate in the interior region, and $d\Omega^2=d\theta^2+\sin^2\theta d\phi^2$ is the metric of the unit 2-sphere in terms of the polar and azimuthal angles, $\theta$ and $\phi$. It is straightforward to realize that the instants when $p_b=0$, or $\cos^2(\delta_bb)=1$ according to Eq. \eqref{pb}, correspond to Killing horizons related to the Killing vector $\partial_x$, beyond which $p_b$ becomes imaginary \cite{AOS,AOS2}. In previous works, the section corresponding to $\cos(\delta_bb)=1$ has been interpreted as a black horizon, whereas the one corresponding to $\cos(\delta_bb)=-1$ has been related to a white horizon instead\footnote{This interpretation is a matter of convention. Without loss of generality, we follow Refs. \cite{AOS,AOS2} and assign times smaller than the value corresponding to the black horizon to the interior region. Further specifications will be made when discussing the matching of the interior and exterior solutions in Sec. \ref{subsec:Exterior}.}. With a general choice of integration constants, these horizons, which constitute the boundaries of the interior region, are easily seen to be located at
\begin{align}
t^{\rm BH}=\dfrac{2}{b_o}\left(\arctanh b_o^{-1}-C_2\right),\quad t^{\rm WH}=-\dfrac{2}{b_o}\left(\arctanh b_o^{-1}+C_2\right)=t^{\rm BH}-\dfrac{4}{b_o}\arctanh b_o^{-1}.
\end{align}
Therefore, the choice of $C_2$ fixes the position of either the black or the white horizon (the other being immediately fixed as well). For instance, we can express $C_2$ as
\begin{align}
C_2=\arctanh b_o^{-1}-\dfrac{1}{2}b_ot^{\rm BH}.
\end{align}
As regards the integration constant $C_1$, let us begin by commenting that $p_c$ (its first derivative with respect to time being proportional to the expansion of null vector fields normal to the metric 2-spheres of constant $t$ and $x$) plays a key role in the analysis of the causal structure of the effective interior geometry \cite{AOS,AOS2}. Along any dynamical trajectory, there is (at most) one instant where $p_c'$ vanishes (in other words, where $p_c$ attains its minimum value). This instant has been argued to correspond to a hypersurface that is the non-singular effective analogue of the classical singularity and that separates a trapped region from an anti-trapped one: the so-called \emph{transition surface} $\mathcal{T}$. A trivial calculation shows that it is located at 
\begin{align}
t^{\mathcal{T}}=\dfrac{1}{2}\ln |C_1|.
\end{align}
Hence, $C_1$ (or, rather, its absolute value) can be understood as determining where this transition surface lies. Notice that we can choose our origin of time to coincide e.g. with the black horizon, or with the transition surface, fixing in this way either the constant $C_2$ or $|C_1|$, but not both of them. 

Before concluding this section, let us briefly comment on the choice of integration constants that lead to the AOS model as presented in Refs. \cite{AOS,AOS2}. In those works, very concrete choices of the integration constants $C_1$ and $C_2$ were made. On the one hand, the black horizon was identified with the hypersurface $t=0$, a possibility that we contemplated above. This fixes $C_2^{\rm AOS}=\arctanh b_o^{-1}$. On the other hand, the remaining integration constant was fixed as $C_1^{\rm AOS}=\gamma L_o\delta_c/(8|m|)$. This choice reproduces the classical relation between the black horizon area and the Hamiltonian mass when $|m|\gg L_o\delta_c$. As we will comment in the next section, restricting the analysis to this value of the integration constant $C_1$ is excessive, because the requirement of a certain classical limit should at most fix its dominant contribution when the polymerization parameters tend to zero. Note that the choice made in the original AOS solution is such that $8|\bar p_c^0| C_1^{\rm AOS}=(\gamma L_o\delta_c)^2$. Thus, any choice that fails to satisfy this relation immediately lies beyond the scope of previous works. This leaves a door open to the existence of unexplored behaviors in the effective geometry. We will comment on this issue in Sec. \ref{subsec:Exterior}, where we will discuss the extension of the model to the exterior region and examine the matching with the interior one. 

\subsection{Exterior region and matching at the horizon}\label{subsec:Exterior}

As we have seen, there exist two instants of time along any dynamical trajectory in the interior region where $p_b$ vanishes and beyond which this variable becomes imaginary. Their spacetime counterparts have been argued to be horizons that play the role of boundaries of the interior geometry. Their existence raises the question of the extension of the formalism beyond those horizons. This was discussed in Refs. \cite{AOS,AOS2}, introducing an approach that circumvents the problems to foliate the exterior region with homogeneous hypersurfaces, which become \emph{timelike} in fact. One can then carry out a treatment analogous to that developed for the interior region, except that the Hamiltonian constraint on the canonical variables defined outside (which we will call $\tilde b$, $\tilde c$, $\tilde p_b$, and $\tilde p_c$) generates an evolution {\sl in the radial direction} \cite{AOS2}. In practice, the counterpart of the dynamical equations \eqref{c'}-\eqref{pb'} in the exterior turns out to be obtained with the following replacements: $b\to i\tilde b$, $c\to\tilde c$, $p_b\to i\tilde p_b$, and $p_c\to\tilde p_c$ (we encourage the reader to consult Ref. \cite{AOS2} for further details). 

The corresponding solutions for the exterior region can be obtained in a manner that is totally analogous to that discussed in the previous subsection, resulting in\footnote{Even though the interior and exterior regions are in general described in terms of different coordinate patches, for simplicity we do not change the notation for their coordinates. It should be clear from the discussion which region we are referring to at any time.}
\begin{align}
\tan \left[\dfrac{\tilde \delta_c\tilde c(t)}{2}\right]&=\tilde C_1 e^{-2t}=\textrm{sgn}(\tilde C_1)\tilde x_c(t),\qquad \tilde p_c( t)=\dfrac{1}{2}\textrm{sgn}(\tilde C_1)\gamma L_o\tilde \delta_c\tilde m\left[\tilde x_c(t)+\dfrac{1}{\tilde x_c(t)}\right],\\
\cosh \left[\tilde \delta_b\tilde b(t)\right]&=\tilde b_o\tanh\left(\dfrac{1}{2}\tilde b_o  t+\tilde C_2\right),\qquad  \tilde p_b(\tilde t)=-2\dfrac{\tilde \delta_b}{\tilde \delta_c}\dfrac{\sinh\left[\tilde\delta_b\tilde b(t)\right]}{\gamma^2\tilde \delta_b^2-\sinh^2\left[\tilde \delta_b\tilde b( t)\right]}\tilde p_c( t)\sin\left[\tilde \delta_c\tilde c( t)\right],
\end{align}
where $\tilde C_1$, $\tilde C_2$, and $\tilde m$ are real integration constants ($\tilde C_1$ and $\tilde m$ being non-zero), and $\tilde x_c(t)=|\tilde C_1|e^{-2t}$. Furthermore, $\tilde \delta_b$ and $\tilde \delta_c$ are two polymerization parameters, that might differ from their interior counterparts, and $\tilde b_o=\sqrt{1+\gamma^2\tilde \delta_b^2}$. 

A priori, the interior and exterior solutions are independent. Nevertheless, in order to provide them with a physical interpretation and construct a smooth ``black hole" geometry, one performs the following matching. Consider the two values $t^{\rm BH}$ and $\tilde t^{\rm BH}$ such that $p_b(t^{\rm BH})=0=\tilde p_b(\tilde t^{\rm BH})$. By setting them equal to each other, which in particular leads to $\tilde C_2=\arctanh \tilde b_o^{-1}-\frac{1}{2}\tilde b_o t^{\rm BH}$, we may match the solutions at the hypersurface $t=t^{\rm BH}$. This hypersurface separates the interior region, corresponding to $t^{\rm WH}<t<t^{\rm BH}$, from the black hole exterior, corresponding to $t>t^{\rm BH}$. 

Let us illustrate the main properties of the geometry resulting from the proposed matching across the black horizon. This requires a change of coordinates to remove the apparent singularity that arises owing to our use of standard Schwarzschild coordinates. Let us focus, for instance, on the effective metric in the exterior region, which can be written as $ds^2_{{\rm ext},2}=-\tilde f_1(t)dx^2+\tilde f_2(t)dt^2$ with 
\begin{align}
\tilde f_1=\dfrac{\tilde p_b^2}{L_o^2|\tilde p_c|},\qquad \tilde f_2=\dfrac{\gamma^2\tilde \delta_b^2|\tilde p_c|}{\sinh^2(\tilde \delta_b\tilde b)},
\end{align}
and where we have omitted the angular part of the metric, because it does not play an important role in the analysis of the near-horizon geometry. Introducing the ingoing Eddington-Finkelstein coordinate $v$, we can reexpress the exterior metric as $ds^2_{{\rm ext},2}=-\tilde f_1 dv^2+2(\tilde f_1\tilde f_2)^{1/2}dvdt$. Although $\tilde f_1$ vanishes at the horizon, this metric remains well defined as long as its determinant differs from zero, condition that amounts to require that the product $\tilde f_1\tilde f_2$ be strictly positive in a neighborhood of the black horizon. It is easy to check that $\tilde f_1\tilde f_2$ equals $4\tilde m^2$ at this horizon, a fact that ensures the non-degeneracy of the metric if $\tilde m\neq 0$ (property that holds by definition). Additionally, the integration constants in the exterior region can be chosen in such a way that the effective metric components are at least $\mathcal{C}^2$ (i.e., they possess continuous second derivatives). This can be seen with a direct calculation. For the exterior metric, the derivatives at the black horizon are given by
\begin{align}
\tilde f_1'\to \dfrac{8|\tilde m|}{\gamma L_o\tilde \delta_c}\dfrac{\tilde x_c\left(t^{\rm BH}\right)}{1+\tilde x_c^2\left(t^{\rm BH}\right)},\qquad &\tilde f_1''\to \dfrac{8|\tilde m|}{\gamma L_o\tilde \delta_c}\dfrac{\tilde x_c\left(t^{\rm BH}\right)}{1+\tilde x_c^2\left(t^{\rm BH}\right)}\left[\dfrac{1}{2}\gamma^2\tilde \delta_b^2+3-4\dfrac{1-\tilde x_c^2\left(t^{\rm BH}\right)}{1+\tilde x_c^2\left(t^{\rm BH}\right)}\right],\\
\left(\sqrt{\tilde f_1 \tilde f_2}\right)'\to 2|\tilde m|,\qquad &\left(\sqrt{\tilde f_1 \tilde f_2}\right)''\to 2|\tilde m|\left(1+\dfrac{1}{2}\gamma^2\tilde\delta_b^2\right).
\end{align}
The corresponding expressions for the interior metric are identical except for the replacement of the tilded constants and polymerization parameters with untilded ones. Therefore, we obtain a metric that is (at least) twice continuously differentiable and non-degenerate by setting $\delta_b=\tilde \delta_b$, $\delta_c=\tilde \delta_c$, $|C_1|=|\tilde C_1|$, and $|\tilde m|=|m|\neq 0$. 

A similar matching can be carried out between another exterior solution, defined for $t<t^{\rm WH}$, and the interior geometry at the white horizon. This yields for the integration constant $\tilde{C}_2$ the value $\arctanh \tilde b_o^{-1}-\frac{1}{2}\tilde b_o t^{\rm WH}$. The remaining integration constants are again fixed by requiring that the geometry be non-degenerate and at least twice continuously differentiable.

In summary,  a general choice of integration constants for the AOS model leads to a smooth and non-degenerate effective geometry described by:
\begin{itemize}
\item An exterior solution for $t>t^{\rm BH}$, determined by two polymerization parameters and the three integration constants $|\tilde{m}|=|m|$, $|\tilde{C_1}|=|C_1|$, and $\tilde C_2=\arctanh  b_o^{-1}-\frac{1}{2} b_o t^{\rm BH}$.
\item An interior solution for $t^{\rm WH}<t<t^{\rm BH}$, labeled by two polymerization parameters and the three integration constants $|m|$, $|C_1|$, and $C_2=\arctanh b_o^{-1}-\frac{1}{2}b_ot^{\rm BH}$.
\item An exterior solution for $t<t^{\rm BH}-4b_o^{-1}\arctanh b_o^{-1}$, described by two polymerization parameters and the three integration constants $|\tilde{m}|=|m|$, $|\tilde{C_1}|=|C_1|$, and $\tilde C_2=3\arctanh  b_o^{-1}-\frac{1}{2}b_o t^{\rm BH}$.
\end{itemize}

The two polymerization parameters can be fixed by minimum area arguments, as it was done in Refs. \cite{AOS,AOS2}, where the authors gave an expression of $\delta_b$ and $\delta_c$ in terms of $|m|$ for sufficiently massive solutions. The other constants, as we have seen, have different interpretations. Firstly, $m$ is a constant of motion related to the mass of the black hole under consideration (see Sec. \ref{sec:AsympMass} for a more detailed view on this matter). Secondly, $C_2$ is related to the freedom to select the instant of time corresponding to, e.g., the black horizon. Thirdly, $C_1$ is related to the position of the only local minimum of the triad variables $p_c$ and $\tilde p_c$. At this stage of the discussion, we can provide an argument to constrain the physically admissible values of the integration constant $C_1$. As the form of the effective metric \eqref{metric} suggests, $|p_c|$ (and similarly $|\tilde p_c|$) is related to the physical area of the metric 2-spheres defined by constant values of $t$ and $x$. For finite values of the polymerization parameters, $|p_c|$ is found to be bounded from below, reaching at most a single minimum in the interior region for a certain value of $t$. If this minimum does not occur, the fact that $|C_1|=|\tilde C_1|$ implies that $|\tilde{p}_c|$ will reach a minimum in one (and only one) of the exterior regions. Notice that, as it stands, $\ln|C_1|$ can adopt any real value and, therefore, so can the position of the transition surface, $t^{\mathcal{T}}$. If, following physical criteria, we want to localize the surface with minimum area of the metric 2-spheres in the interior region, we need to restrict the admissible values of $|C_1|$. This requirement directly translates into the condition
\begin{align}
2t^{\rm BH}-\dfrac{8}{b_o}\arctanh b_o^{-1}<\ln|C_1|<2t^{\rm BH},\label{bounds}
\end{align}
which provides at least an upper bound on the value of $|C_1| \textrm{exp}(-2t^{\rm BH})$. 

Further restrictions can be derived upon inspecting the classical limit of the model and requiring that it is compatible with the predictions of general relativity, at least for large black holes. If we want to identify the square of the Schwarzschild radius with $|p_c(t^{\rm BH})|$ in the limit $|m|\gg L_o\delta_c$, it is straightforward to realize that we must have $|C_1|\textrm{exp}(-2t^{\rm BH})\sim \frac{\gamma L_o\delta_c}{8|m|}$ at leading order\footnote{As we show in Sec. \ref{sec:AsympMass}, the ADM mass of the spacetime for any solution of the AOS equations is exactly given by $|m|/G$. It follows that the Schwarzschild radius, standardly defined as the product of $2G$ and the ADM mass, is simply equal to $2|m|$.}. This dominant term reduces to $C_1^{\rm AOS}$ when we set $t^{\rm BH}=0$. However, we can choose a much more general $|C_1|$, allowing for corrective terms in the limit of very large masses,
\begin{align}\label{physicalc1}
|C_1|e^{-2t^{\rm BH}}=\dfrac{\gamma L_o \delta_c}{8|m|}+o\left(\dfrac{L_o\delta_c}{|m|}\right),
\end{align}
where $o(\cdot\,)$ denotes terms that are subdominant with respect to the quantity in parenthesis. Moreover, in general these corrections might even dominate away from the large mass regime as long as the bounds \eqref{bounds} are satisfied. 

%%%%
\section{Thermodynamical properties}\label{sec:Thermodynamics}

Once we have shown that the interior and exterior solutions can be satisfactorily matched at the black and white horizons, we can proceed to examine some of the physical properties of the resulting model. In this section, we will focus on the thermodynamical properties of the black horizon, carrying out an analysis similar to that of Ref. \cite{AO}.

The vector field $\partial_x$ is a Killing vector of the metric, with a norm in the exterior region equal to $-\tilde f_1$. Therefore, $\partial_x$ is timelike away from the horizon and becomes null only at the very horizon. This property characterizes the hypersurface defined by $t=t^{\rm BH}$ as a Killing horizon. To investigate the associated thermodynamical behavior, let us perform a Wick rotation $x\to x_E=-ix$ to the exterior spacetime metric. We obtain
\begin{align}
ds^2_{\rm ext}=\tilde f_1(t)dx_E^2+\tilde f_2(t)dt^2+|\tilde p_c|d\Omega^2.
\end{align}
Since this metric has Riemannian signature and the norm of $\partial_{x_E}$ vanishes at the horizon, we conclude that this Killing vector field vanishes there. As a consequence, in a neighborhood of the horizon, $\partial_{x_E}$ resembles the generator of a rotation in the $t$-$x_E$ plane. Ignoring the angular part of the metric and performing the change of variable $R=[\tilde f_1(t)]^{1/2}$, we immediately get
\begin{align}
ds^2_{{\rm ext}, 2}=R^2dx_E^2+\dfrac{4\tilde f_1\tilde f_2}{(\tilde f_1')^2}dR^2.
\end{align}
To make sure that this metric does not present a conical singularity at the horizon, i.e. at $R=0$, let us examine the ratio between the physical length $L$ of a circumference of coordinate radius $R=\delta R$ and its physical radius $r$ as $\delta R$ tends to zero. Assuming periodicity of the coordinate $x_E$ with period $\mathcal{P}$, we obtain\footnote{We assume that $\tilde f_1'/(\tilde f_1\tilde f_2)^{1/2}$ is approximately constant in the interval of integration that provides the physical radius when $\delta R\to 0$.} 
\begin{align}
\lim_{\delta R\to 0}\dfrac{L}{r}=\lim_{t\to t^{\rm BH}}\dfrac{\mathcal{P}\tilde f_1'(t)}{2\sqrt{\tilde f_1(t)\tilde f_2(t)}}.
\end{align}
It is straightforward to check that this limit is well defined and non-zero, since $\lim_{t\to t^{\rm BH}}\tilde f_1\tilde f_2=4 m^2>0$ and $\lim_{t\to t^{\rm BH}}\tilde f_1'\neq 0$ for $m\neq 0$, as we saw in Sec. \ref{subsec:Exterior}. Employing the explicit formulas of that section and requiring that the limit of the considered ratio be $2\pi$, we find that the period $\mathcal{P}$ must adopt a very concrete, non-zero value for the Wick-rotated exterior metric to be regular at the horizon. Owing to this periodicity, test quantum fields that propagate on the effective exterior geometry display features of a thermal state with temperature \cite{Hawkingp}
\begin{align}\label{genTH}
T_H=\dfrac{1}{k_B\mathcal{P}}=\dfrac{1}{8\pi k_B |m|}\left\{\dfrac{8|m| \tilde x_c(t^{\rm BH})}{\gamma L_o \delta_c [1+\tilde x_c^2(t^{\rm BH})]}\right\},
\end{align}
where $k_B$ is the Boltzmann constant. With the choice of integration constants made in Refs. \cite{AOS,AOS2}, one recovers
\begin{align}\label{THAOS}
T_H^{\rm AOS}=\dfrac{1}{8\pi k_B |m| (1+\epsilon_m)},
\end{align}
where $\epsilon_m=\gamma^2L_o^2\delta_c^2/(64m^2)$, which is the result obtained in Ref. \cite{AO}. Note, however, that a more general choice of integration constants does affect the temperature of the quantum fields on the effective geometry of the model, resulting in a relative variation  
\begin{align}\label{reldifTH}
\dfrac{|T_H-T_H^{\rm AOS}|}{T_H^{\rm AOS}}=\dfrac{\tilde{x}_c(t^{\rm BH})}{1+\tilde{x}_c^2(t^{\rm BH})}(\epsilon_m^{1/2}+\epsilon_m^{-1/2})-1.
\end{align}
In the limit $L_o\delta_c\ll|m|$ considered at the end of Sec. \ref{sec:GeneralSolution}, a relativistic behavior imposes that $\tilde{x}_c(t^{\rm BH})$ and $\epsilon_m^{1/2}$ be small and approach each other, as illustrated by Eq. \eqref{physicalc1}. A good estimation of the relative variation \eqref{reldifTH} in the mentioned regime is then given by $\epsilon_m^{-1/2}\tilde{x}_c(t^{\rm BH})-1$, i.e. by the relative change in $\tilde{x}_c(t^{\rm BH})$ with respect to $\epsilon_m^{1/2}$. The value adopted by this quantity naturally depends on how the integration constants are fixed. Although the resulting contribution should generally be small with respect to the unit in the discussed limit, we emphasize that it may be larger than $\epsilon_m$ and, thus, larger than the corrections to the general relativistic Hawking temperature arising in the original AOS solution [see Eq. \eqref{THAOS}].

We can see in a straightforward manner that the temperature \eqref{genTH} depends on the relation between the physical area of the black horizon and the mass $|m|$, relation that depends in turn on the integration constants. Let us call this area $4\pi r_S^2$, with $r_S^2=|\tilde p_c(t^{\rm BH})|$. Direct comparison with the Schwarzschild metric in general relativity shows that $ r_S$ must tend to the value $2|m|$ in the classical limit, which is the reason behind the notation chosen to denote this quantity. Solving a simple second order polynomial equation, we can rewrite $\tilde x_c(t^{\rm BH})$, which appears in $|\tilde p_c(t^{\rm BH})|$, in terms of $ r_S^2$. Out of the two possible solutions, which are related by means of a simple inversion, the one with the appropriate classical limit is
\begin{align}
\tilde x_c(t^{\rm BH})=|C_1|e^{-2t^{\rm BH}} =\dfrac{r_S^2}{4m^2}\dfrac{4|m|}{\gamma L_o \delta_c}\left[1-\sqrt{1-\left(\dfrac{4m^2}{r_S^2}\right)^2\left(\dfrac{\gamma L_o \delta_c}{4m}\right)^2}\right].
\end{align}
Using this expression, the Hawking temperature can be rewritten as
\begin{align}
T_H=\dfrac{\hbar}{8\pi k_B|m|}\dfrac{4m^2}{ r_S^2}.
\end{align}
Thus, the quantum effects are encoded in the quotient $4m^2/ r_S^2$, which we do not force to be equal to the unit. The fact that this quotient can be, in general, different from one for finite values of the mass is a consequence of the fact that, generally, $|m|/G$ need not agree with $M_{\rm H}=r_S/2G$, which we will call the horizon mass. Indeed, the precise relation between these two masses is a function of $C_1$ and $C_2$ (or $t^{\rm BH}$),
\begin{align}
M_{\rm H}=\sqrt{\dfrac{\gamma L_o\delta_c}{8|m|}\left[|C_1|e^{-2t^{\rm BH}}+\dfrac{e^{2t^{\rm BH}}}{|C_1|}\right]}\dfrac{|m|}{G}.\label{MH}
\end{align}
This does not mean that $|m|/G$ does not coincide with other notions of mass for the considered solutions, such as the ADM mass, as we will see in Sec. \ref{sec:AsympMass}. Rather, the situation that we have found shows that alternative definitions of the mass that are known to coincide in general relativity need not lead to the same result in more general scenarios, e.g., when quantum gravitational effects are taken into account. Bearing in mind our comments below Eq. \eqref{reldifTH}, one can convince oneself that the departure from the value of the horizon mass in the original AOS model is intimately related to the previously discussed deviation in the Hawking temperature. Actually, a straightforward calculation shows that, in the aforementioned limit $L_o\delta_c\ll |m|$, the dominant term in $GM_H/|m|-1$ is either of order $\epsilon_m$ or of the same order as the relative difference between temperatures given in Eq. \eqref{reldifTH}, whichever one is larger in the regime under consideration for a given choice of admissible integration constants.

%%%%
\section{Asymptotic behavior and notions of mass}\label{sec:AsympMass}

Let us turn now our attention to the analysis of the effects of a general choice of integration constants on the asymptotic behavior of the geometry and on different definitions of the mass.

The asymptotic behavior of the original AOS model was studied in Refs. \cite{AO,Bouhm}, with the conclusion that the effective exterior geometry is asymptotically flat only \emph{in an elementary sense}. Indeed, the exterior metric approaches a flat metric at spatial infinity, as proven in Ref. \cite{AO}, but it does not do so at a sufficiently fast rate so as to satisfy certain conditions on the fall-off of its derivatives \cite{AO,Bouhm}. The result for our general setting turns out to be analogous. After recasting the effective metric in a form that is better suited to the analysis at spatial infinity, we can perform an asymptotic expansion to identify the dominant terms in each metric component. As is the case of the original AOS solution, we find that, provided that the polymerization parameters are different from zero, the time component of the metric diverges at spatial infinity. Let us recall that, given the physical interpretation of the polymerization parameters, this implies that there appears a divergence in the metric if there are quantum effects in the system, no matter how small. Following the same procedure as in Ref. \cite{AO}, we can reabsorb the divergent factor through an appropriate time redefinition. After this change of time, the effective metric at spatial infinity manifestly approaches a flat metric. However, this change also introduces certain terms that are relevant for the derivatives of the metric tensor, which then fails to satisfy the conditions necessary for a standard notion of asymptotic flatness \cite{flatness,wald}. 

On the other hand, a concept of great relevance for black hole spacetimes is their mass. In general relativity, there exist several proposals to define the mass associated with a certain geometry, many of which involve its asymptotic regions. This is the case of, e.g., the ADM mass. For the sake of comparison, it is also interesting to consider other notions of mass for black hole geometries. Another definition that has been employed in recent analyses involves the Ricci tensor of the spatial metric (for more details, see Ref. \cite{33}). It is well known that, in the case of the Schwarzschild metric (when the geometry under consideration is asymptotically flat in the standard sense), both definitions of the mass yield the same result. However, this need not be true in scenarios such as the one in hand, where effects of quantum origin have been captured through the introduction of polymerization parameters. 

Let us consider the ADM mass first, which is defined in terms of the spatial part $\tilde q_{ab}$ of the effective metric in the exterior. In order to express it in a convenient form, we introduce the change of coordinates $\tilde r=\tilde r_S e^{(t-t^{\rm BH})}$, where $\tilde r_S^2=\gamma L_o\delta_c |m|e^{2t^{\rm BH}}/(2|C_1|)$. Note that, for choices of $|C_1|$ that lead to solutions with an acceptable classical behavior in the limit of large masses, this constant coincides to leading order with the quantity $r_S^2$ introduced in the previous section [see Eq. \eqref{physicalc1} and the paragraph below Eq. \eqref{THAOS}]. In terms of the new radial coordinate $\tilde r$ and the parameter $\epsilon=b_o-1$, the spatial part of the exterior metric can be written as
\begin{align}
\tilde q_{ab}dx^adx^b=\dfrac{\left(2+\epsilon+\epsilon \,{\xi^{1+\epsilon}}\right)^2\left(1+|C_1|^2{e}^{-4t^{\rm BH}}{\xi}^{4} \right)}{\left(1- {\xi}^{1+\epsilon}\right)\left[(2+\epsilon)^2-\epsilon^2 {\xi}^{1+\epsilon}\right]}d\tilde r^2+\tilde r^2\left(1+|C_1|^2{e}^{-4t^{\rm BH}}{\xi}^{4} \right) d\Omega^2,
\end{align}
where $\{x^a\}_{a=1,2,3}=\{\tilde r,\theta,\phi\}$ and $\xi={\tilde r_S}/{\tilde r} $. On the other hand, the square of the lapse function in the exterior region $\tilde N$ is given by
\begin{align}
\tilde N^2=\dfrac{4m^2}{\tilde r_S^2}\dfrac{\left(1- {\xi}^{1+\epsilon}\right)\left[(2+\epsilon)^2-\epsilon^2 {\xi}^{1+\epsilon}\right]\left(2+\epsilon+\epsilon \, {\xi}^{1+\epsilon}\right)^2}{16(1+\epsilon)^4 \left(1+|C_1|^2{e}^{-4t^{\rm BH}}{\xi}^{4}\right)}{\xi}^{-2\epsilon }.
\end{align}
With these expressions, we can calculate the ADM mass, which can be defined as \cite{ADMT}
\begin{align}
M_{\rm ADM}=\dfrac{1}{16\pi G}\lim_{\tilde r\to\infty}\oint_{\tilde r}dS_d \sqrt{{\rm det}(\tilde q)}\tilde N \tilde q^{ac}\tilde q^{bd}(\partial_c\tilde q_{ba}-\partial_b\tilde q_{ca}), 
\end{align}
where the integration must be performed over a 2-sphere of constant radius $\tilde r$, and the partial derivatives must be taken with respect to the Cartesian coordinates $\{y^a\}$ of the asymptotic flat metric to which the spatial effective metric $\tilde q_{ab}$ tends at spatial infinity. Besides, $y^a dS_a =\tilde{r}d^2 V$, where $d^2V$ is the area element of the considered 2-sphere. A direct computation finally yields 
\begin{align}
M_{\rm ADM}=\dfrac{|m|}{G}\dfrac{{\tilde r_S}^{1+\epsilon}(2+\epsilon)^2 }{4(1+\epsilon)^2}\lim_{\tilde r\to\infty} \frac{\tilde r^2 \tilde q_{\tilde r\tilde r}(\tilde r)- \tilde r \partial_{\tilde r}\tilde q_{\theta \theta}(\tilde r)+\tilde q_{\theta \theta}(\tilde r)}{\tilde r^{3+\epsilon}} =\dfrac{|m|}{G}.
\end{align}
The last equality holds for $\epsilon <3$, whereas for $\epsilon >3$ the asymptotic limit diverges. Note that $\epsilon <3$ by definition for macroscopic black holes \cite{AOS2}.  Moreover, this bound must be satisfied if the quantum gravitational effects are reasonably small. Focusing on situations of this type, we conclude that the parameter that we have been calling the Hamiltonian mass is indeed proportional to the ADM mass, $|m|=GM_{\rm ADM}$. Notice that this result is identical to the one found in Ref. \cite{AO}. This is an immediate consequence of the fact that the ADM mass is independent of the choice of the integration constants $C_1$ and $C_2$. We recall that this was not the case when we defined the horizon mass by measuring the analog of the Schwarzschild radius, resulting into a dependence on the value of $|C_1|\textrm{exp}(-2t^{\rm BH})$ [see Eq. \eqref{MH}].

Let us study now the second definition of the mass that we mentioned above, namely the one constructed with the Ricci tensor of the spatial metric $\tilde{\mathcal{R}}^{(3)}_{ab}$. This mass is defined as
\begin{align}
M_{\rm Ricci}=\dfrac{1}{8\pi G}\lim_{\tilde r\to\infty}\oint_{\tilde r}d^2V\tilde r\tilde N\tilde{\mathcal{R}}^{(3)}_{ab}\hat r^a\hat r^b,
\end{align}
where the integration is again over a 2-sphere of constant radius $\tilde r$, and $\hat r^a$ is a unit radial vector field. Since there is no integration in the radial direction, it suffices to compute the dominant term of the $\tilde r\tilde r$-component of the Ricci tensor, which is a simple task given that the spatial effective metric $\tilde q_{ab}$ is diagonal. With these considerations, it is easy to show that
\begin{align}
M_{\rm Ricci}= (1+\epsilon) \dfrac{|m|}{G}=(1+\epsilon)M_{\rm ADM}=(1+\epsilon)\left\{\dfrac{\gamma L_o\delta_c}{8|m|}\left[|C_1|e^{-2t^{\rm BH}}+\dfrac{e^{2t^{\rm BH}}}{|C_1|}\right]\right\}^{-1/2}M_{\rm H}.
\end{align}
Hence, the three studied definitions of mass, which are known to coincide when applied to a classical Schwarzschild spacetime, differ in our solutions. In the case of the ADM and Ricci masses, the difference is given by a global factor that incorporates quantum gravitational effects (it is proportional to $\epsilon$) and vanishes when these effects disappear. Again, this is the same result that was found in Ref. \cite{AO}. It is worth remarking that these notions of mass continue to be well defined (at least for small quantum effects), even in the absence of standard asymptotic flatness, and the resulting values only disagree slightly for massive black holes. The case of the horizon mass is somewhat different, since its value depends on the choice of the integration constants $C_1$ and $C_2$ (or rather, on the combination $|C_1|e^{-2t_{\rm BH}}$). 

%%%%
\section{Conclusions}\label{sec:Conclusions}

The AOS model \cite{AOS,AOS2} to describe Schwarzschild black holes within the context of LQC has appealing features that have attracted considerable attention. The central singularity that appears in general relativity is replaced with a transition surface connecting a trapped region with an anti-trapped one. These regions are bounded by black and white horizons, respectively, beyond which the formalism can be extended to account for the entire Kruskal spacetime, leading to an effective geometry that is smooth and non-singular. Its curvature invariants are bounded from above by quantities that are independent of the black hole mass. Given these interesting properties, the next logical step would be to proceed to the quantization of the black hole spacetime along the lines of LQC. In order to do this, it is fundamental to have a neat understanding of the space of solutions of the dynamical equations. Nonetheless, in Refs. \cite{AOS,AOS2} the constants that appear in the integration of the equations of motion were chosen in a very particular way, and an investigation of the full space of solutions is absent in the literature. The aim of this article is to fill this gap and explore the consequences of the most general choice of integration constants that can be allowed. This study provides a firm groundwork to construct a fully quantum description of Schwarzschild black holes within the LQC framework. In addition, it also clarifies how the choice of the integration constants affect physical features of the model such as its thermodynamical properties or different definitions of the black hole mass.

In Sec. \ref{sec:GeneralSolution}, we have discussed the general solution to the dynamical equations, both in the interior and the exterior regions (see Secs. \ref{subsec:Interior} and \ref{subsec:Exterior}, respectively). Each of these solutions depends on three integration constants and two positive polymerization parameters (that are functions of the mass with an asymptotic behavior that is fixed by minimum area conditions). We have then addressed the matching of the interior and exterior solutions at the horizons, requiring a non-degenerate and sufficiently smooth geometry. In this way, we have related the integration constants inside and outside the horizon. As a consequence, the determination of a solution, valid in the interior as well as in the exterior, amounts to the choice of three constants (other than the polymerization parameters), which are related to three physical quantities, namely, the Hamiltonian mass, the position of the black horizon, and the position of the transition surface replacing the singularity. We have commented on the values of the integration constants that reproduce the results of Refs. \cite{AOS,AOS2}, as well as on how to restrict the physically acceptable values of one of the constants by demanding that the transition surface belong to the interior region. In Sec. \ref{sec:Thermodynamics}, we have computed the Hawking temperature associated with the black horizon, which we have found to depend on the choice of integration constants. Indeed, we have shown that the temperature depends on the relation between the black horizon area and the Hamiltonian mass. This relation does not need to be the same as in general relativity, so that the choice of integration constants leaves an imprint in the thermodynamics of the model. In Sec. \ref{sec:AsympMass}, we have seen that the asymptotic structure of the model is similar to that studied in Ref. \cite{AO}. Finally, we have examined different definitions for the mass of the spacetime that are known to coincide in the Schwarzschild case. For instance, we have computed the ADM and Ricci masses, which turn out to be independent of the choice of the position of the horizon and of the transition surface. We have shown that for any solution they just differ by a term that vanishes in the absence of quantum effects, just like it happened in Ref. \cite{AO}.

Having gained this knowledge of the space of solutions of the AOS dynamics, it would be especially interesting to explore whether the studied dynamical equations can be reconciled with a canonical treatment where the polymerization parameters are fixed as constants of motion, providing a clear connection between these equations and the proposed effective Hamiltonian (therefore surpassing the problems described in Ref. \cite{N}). With this goal in mind, the way forward involves a careful examination of the procedures presented in Ref. \cite{AOS}, considering in particular a suitable extension of the phase space and its subsequent reduction. These issues will be addressed in a future investigation. 

%%%%

\acknowledgments
	
This work has been supported by Project. No. MICINN PID2020-118159GB-C41. B. Elizaga Navascués acknowledges financial support from the Standard program of JSPS Postdoctoral Fellowships for Research in Japan. A. García-Quismondo acknowledges that the project that gave rise to these results received the support of a fellowship from ``la Caixa'' Foundation (ID 100010434). The fellowship code is LCF/BQ/DR19/11740028.

\end{document}